\documentclass{appolb}

\begin{document}

\title{STATISTICAL AND DYNAMICAL MEASURES OF SIMPLE IRREVERSIBLE
PROCESSES}
\author{Mariusz Soza\'nski%
\thanks{Electronic address: Mariusz.Sozanski@if.pw.edu.pl}, Jan \.Zebrowski
\address{Faculty of Physics, Warsaw University of Technology, Koszykowa 75,
00-662~Warsaw, Poland}}
\maketitle

\begin{abstract}
A simple model of an irreversible process is introduced. The
equation of iterations in the model includes a noise generation
term. We study the properties of the system when the noise
generation term is a stochastic process (\eg~a random
number generator) or a deterministic process (\eg~a chaotic map).
We compare the time series obtained from the above implementations
of the model by use of statistical methods (such as Detrended
Fluctuation Analysis). The conclusion is that using statistical
methods the two versions of the model are indistinguishable. The
advantage of this observation is that we may calculate the
Lyapunov exponent for the model. As a result we obtain an equation
relating the DFA exponents (a statistical measure) with the
Lyapunov exponent for such models. On the other hand, typical
statistical properties can also be calculated, as for example the
diffusion coefficient for a particle, which movement is defined by
the above model.
\end{abstract}

\pagestyle{plain}
\newcount\eLiNe\eLiNe=\inputlineno\advance\eLiNe by -1


\section{Introduction}

Usually deterministic and stochastic analysis are not considered
compatible. On one hand we are often limited to statistical analysis
when dealing with experimental time series.
On the other hand a purely stochastic approach to data analysis may
lead to an erroneous interpretation about the genesis of the
analyzed system.

In this work, we investigate,
whether it is possible to substitute the Lyapunov exponent with a
statistical measure, at least in simple 1-D systems. This may be
valuable in the case when we do not know what is the equation
behind the process we observe. It may also be useful when it is
technically difficult to calculate the Lyapunov
exponent~\cite{Beck93,ZillmerPhD}.

In this paper, we use the so-called general random walk as our
main model~\cite{KlagesPhD,Korabel04,Reichl98}. We analyze the
properties of this model when the noise components of the walk are
generated by a stochastic process or a deterministic process
(\eg~the tent map). Using Detrended Fluctuation
Analysis~(\emph{DFA}) ~\cite{Peng94,Peng95,Talkner00} we compare
the time series obtained from the above implementations of the
model. The conclusion is that both implementations of the model
lead to the same results. As a consequence of this observation, we
calculate the Lyapunov exponent for the deterministic version of
model. The result is an equation relating the DFA exponents (a
statistical measure) with the Lyapunov exponent for our model,
which is practically independent of the method used to generate
noise. Using the Lyapunov exponent we may calculate the diffusion
coefficient for a particle, which movement is defined by the
model. By doing this we also doublecheck that the deterministic
implementation of the model has the same statistical properties as
the purely stochastic one.

\subsection{The method}

Detrended Fluctuation Analysis (\emph{DFA}) has been originally
applied to the DNA walk~\cite{Peng94}. Therefore in general one
needs to define a walk that is related to the input series
analogously~\cite{Peng95}. To do this, the input series $\{X(k)\}$
of length $N$ is integrated after subtracting their average value.
The series $\{y(k)\}$ is then:
\begin{equation}  \label{eq1}
y(k) = \sum_{i=1}^k (X(i) - \left<{X}\right>)\,,
\end{equation}
where $X(i)$ is the $i$-th point of a discrete time series and
$\left<X\right>$ is the average value of the data. Next, the
integrated series is divided into subintervals of equal length
$n$, and for each subinterval a linear least squares fit to the
$y(k)$, denoted $y_n(k)$, is made~(fig.~1). The RMS fluctuation
around the regression line is then given by the equation:
\begin{equation}  \label{eq2}
F(n) = \sqrt{\frac{1}{N}\sum_{k=1}^N [y(k) - y_{n}(k)]^2}\,.
\end{equation}

The dependence of $F$ on the length $n$ is examined via a plot of
log $F(n)$ versus log $n$ (fig.~2). When scaling occurs, the
overall slope of the line in the double-logarithmic scale is equal
to the DFA exponent and denoted by $ \alpha = \log (\e^{-b} F (n))
(\log n)^{-1},$ where $b$ is the intercept of the approximated
trend.

Generally, when a scaling $F(n)\propto n^\alpha$ is observed, the
scaling exponent in the range $0.5<\alpha <1$ indicates positive
long-range power-law correlations (in other words, persistence)
and $ 0<\alpha <0.5$ infers anticorrelations (antipersistence)
\cite {Peng94,Peng95,Talkner00}. $\alpha =1.5$ is obtained for the
Brownian walk. The exponent $\alpha =0.5$ (fig. 2a)  corresponds
to uncorrelated data.

If there are short-range correlations, the slope for low $n$ may
differ from 0.5 but it will approach this value for large $n$. A
crossover occurs when for different ranges of $n$ we observe
different slopes \cite{Peng95}. Crossovers in the linear
dependence of the exponent on window size n have been observed in
detrended fluctuation analysis of complex data (\eg biological
data) and may be an important indicator characterizing the
underlying process \cite{Peng94,Peng95}.

\subsection{The model}

We define the model by the iterative equation:
\begin{eqletters}
\label{eq3}
\begin{equation}
x_{n+1} = x_n + \xi_{n+1} - \xi_{n+1-M}.\label{eq3a}
\end{equation}
Here, $n$ is the iteration index (a natural number), $x_n$ is the
variable value at the $n$-th iteration and $\xi_n$ is the noise
term. The model equation (\ref{eq3a}) is equivalent to:
\begin{equation}
x_n = \sum_{i=0}^{M-1}\xi_{n-i}.\label{eq3b}
\end{equation}
\end{eqletters}
As explicitly written in equation (\ref{eq3b}), our model is a
system with memory, the range of which is defined by parameter
$M$. This is an analogy to the general random walk
process~\cite{KlagesPhD,Korabel04}.

We use two processes to generate the noise $\xi_n$ - a stochastic
and a deterministic one. In the \emph{first} approach we use a
stochastic process that has a uniform distribution with
$\left<\xi\right>=0.5$ and
$\left<\xi_i\xi_k\right>=\delta_{ik}\left<\xi_i^2\right>$.

As a \emph{second} approach we generate noise using chaotic maps.
The following equation defines the tent map (also known as the
symmetric triangular map)~\cite{Beck93,Ott93}:
\begin{equation}\label{tentb}
\xi_{n+1} = f(\xi_n) = 1 - \vert 1 - 2~\xi_n \vert\,.
\end{equation}
It is well-known~\cite{Beck93,Ott93,Schuster} that iterating
eq.~(\ref{tentb}) generates data which is equivalent to
statistically uncorrelated noise with a uniform i.i.d. distribution.
In other words, the natural invariant density~\cite{Beck93,Ott93}
for the tent map~(\ref{tentb}) is equal to $\rho(\xi)=1$.

In this paper we also consider the generation of noise using the
Ulam map~(~\ie~the logistic map at fully developed chaos)~\cite{Beck93}:
\begin{equation}\label{ulam}
\xi_{n+1} = f(\xi_n) = 4\xi_n(1- \xi_n).
\end{equation}
The main difference between the maps (\ref{ulam}) and
(\ref{tentb}) is that the natural invariant density for the Ulam
map does not correspond to the uniform distribution~\cite{Ott93},
instead it is equal to:
\begin{equation}\label{ulamro}
\rho(\xi) = \frac{1}{\pi\sqrt{\xi(1-\xi)}}.
\end{equation}
As we will show further, the differences between the maps
(\ref{tentb}) and (\ref{ulam}) will not affect the results
presented in this paper. To obtain a simple correspondence with
the parameter $x_n$ we define $\xi_n$ as:
\begin{equation}\label{modulo}
\xi_{n+1} =  f(x_{n} mod 1).
\end{equation}
What is important for our study, this approach does not alter the slope
of the chaotic map therefore it preserves the value of the Lyapunov
exponent (though it introduces singularities).

\section{Analysis and comparison of the models}

\bigskip
\subsection{The stochastic model}

First, let us focus on the stochastic version of the model given by
eq.~(\ref{eq3}). As it can be seen in fig. 2 we observe a crossover
for series obtain with parameter $M\not=1$. The point of crossover
is found to be dependent on the parameter $M$. The dependence is a
power-law and for larger $M$ it occurs at larger $n$ (fig.~3).
A similar relation between the correlation range and the crossover
point was assumed in \cite{Peng95} (without proof or
reference to a model) and applied to the analysis of the heart
rate variability series.

We can find this dependence between the crossover point $n_{c}$
and parameter $M$ from numerical data~(fig.~3):

\begin{equation}  \label{logM}
\log n_c = 0.4 \log M + 0.4.
\end{equation}

Therefore, it is possible to obtain the parameter $M$ of the model
(\ref{eq3}) \emph{only by analyzing the time series (experimental
data)}, \eg using the statistical method DFA.

For a pure random walk~($M\to\infty$) it is easy to find the
diffusion coefficient $D$, which in the general case is defined
as~\cite{KlagesPhD,Dorfman99}:
\begin{equation}  \label{diff}
D= \lim_{n\to\infty}\frac{\left<\left(x_n-x_0\right)^2\right>}{2n}
\end{equation}
As $M\to\infty$, eq.~(\ref{eq3}) becomes
\begin{eqletters}
\label{purerw}
\begin{equation}
x_n =\sum_{j=0}^n\xi_j \label{purerwa}
\end{equation}
and, equivalently
\begin{equation}
x_{i+1} = x_i+ \xi_{i+1}.
\label{purerwb}
\end{equation}
\end{eqletters}
Therefore the diffusion coefficient of the pure random walk
(\ref{purerw}) is equal to:
\begin{equation}\label{diffpurerw}
D_s=\frac{\left<\xi^2\right>+\left<\xi\right>^2}{2}=\frac{7}{24}.
\end{equation}

\subsection{The deterministic model}

The dependence of the properties of chaotic maps on the control
parameter is usually described by the Lyapunov exponent, which is
a measure of the memory of the initial conditions. The Lyapunov
exponent for one-dimensional iterated maps is calculated as
~\cite{Ott93,Schuster,Dorfman99}:

\begin{equation}  \label{lap}
\lambda = \lim_{N\to\infty}\lambda(N,x_0) =
\lim_{N\to\infty}\frac{1}{N}
\ln\left|\frac{dF^N(x_0)}{dx_0}\right| =
\lim_{N\to\infty}\frac{1}{N}
\sum_{i=0}^{N-1}\ln\left|F^{\prime}(x_i)\right|,
\end{equation}
where $x_i$ corresponds to the $i$-th iteration of the map
$F(\cdot)$ (in our case defined by eq. (\ref{eq3})) and $x_0$ is
the initial condition. A negative value of $\lambda$ indicates
periodic states, a positive $\lambda$ - chaotic states. In general
the Lyapunov exponent is dependent of the initial condition $x_0$,
but in the case of the map defined by eq.~(\ref{eq3})
$\lambda=\lambda(x_0)$.

Instead of calculating the Lyapunov exponent using the long-time
limit $\lim_{N\rightarrow \infty }\lambda (N)$ as in
eq.~(\ref{lap}), this exponent may be calculated as the average of
all the possible values of the one-step finite-time Lyapunov
exponent $\lambda_i$~\cite{ZillmerPhD, Aurell97}:
\begin{equation}
\lambda =\Big\langle \lambda_i \Big\rangle \  ; \ \  \lambda_i =
\ln |F'(x_i)|. \label{lap1}
\end{equation}

First, we will consider the case of the pure random walk, \ie when
$M\to\infty$. The Lyapunov exponent (eq.~(\ref{lap})) for the
above general walk can be easily found if we note, that the
inversion of the tent map (\ref{tentb}) gives two symmetric
preimages $\xi_n$ for each $\xi_{n+1}$~\cite{Beck93,Schuster}.

To calculate the exponent $\lambda$ in our model we need to find
the product of $2^M$ preimages. However, eq.~(\ref{lap1}) is the
average of one-step Lyapunov exponents $\lambda_i$.
 Due to the symmetry of the
preimages of (\ref{tentb}), there are two equally probable (in the
statistical sense, as $n$ tends to infinity) values of
$\lambda_i$, namely $\ln 1$ and $\ln 3$. Therefore the Lyapunov
exponent for the pure random walk is equal to:
\begin{equation}
\lambda =\frac{1}{2}\ln3.
\label{lappurerw}
\end{equation}
We obtain the same calculation in the case of the Ulam
map~(\ref{ulam}).

Now we will derive the relation between the Lyapunov exponent and
the memory parameter M for the general random walk. At first, this
seems a difficult task, because this time we have to find the
values of the derivatives of all the preimages of $x_{n+1}$ up to
$2^M$ possible values of $x_{n-M+1}$. This step is a little
similar to the construction of the Julia set, in the way that one
needs to find calculate all the preimages values~(compare
\cite{Beck93}, p.104). Yet it is well-known that the Julia set is
quite complex. On the contrary, in this case, after simple
calculations,  a finite sum of a geometrical progression is
obtained, and eq.~(\ref{lap1}) yields:

\begin{eqnarray*}  \label{mylyapunov}
\lambda &=&
\frac{1}{4}\ln\Big[(3+\frac{1}{2^{M-1}})(3-\frac{1}{2^{M-1}})(1+
\frac{1}{2^{M-1}})(1-\frac{1}{2^{M-1}})\Big]  \label{mlap1} \\
\end{eqnarray*}
In the above derivation we use the fact, that in the long-time
limit (\ie when averaging over infinitely many iterations) all
values of the one-step Lyapunov exponents are equally probable. As
we stated earlier, this is due to the symmetry of the two
preimages in both the logistic map and the tent map.
\newline Concluding, we see that the Lyapunov equation depends only on
the parameter $M$ and can be calculated from the equation:
\begin{equation}
\lambda =\frac{1}{4}\ln
\Biggr[\Big(9-2^{2-2M}\Big)\Big(1-2^{2-2M}\Big)\Biggr].
\label{Mlambda}
\end{equation}
Note, that in the limit $M\to\infty$ we obtain
eq.~(\ref{lappurerw}) which was calculated earlier for more
clarity.

The important observation now is that using the DFA method we have
obtained the relation between the the crossover point $n_{c}$ and
$M$~(eq.~(\ref{logM}), see fig. 3). This relation is valid for
both the statistical model of eq.~(\ref{eq3}) and the
deterministic models of eqs.~(\ref {tentb}) and (\ref{ulam}).

This means that we can find the Lyapunov exponent for our model
using only statistical analysis by comparing eqs.~(\ref{Mlambda})
and (\ref{logM}):
\begin{equation}  \label{nlambda}
\lambda = \frac{1}{4}\ln\Biggr[ \Biggr(
1-\frac{1}{2^{2n_c^{5/2}-2}} \Biggr)^2 \Biggr(
1+\frac{1}{2^{n_c^{5/2}-1}} \Biggr)^2 \Biggr] .
\end{equation}

We obtained a quite complicated relation, therefore, to
doublecheck its correctness we show that the Lyapunov exponent
approaches the limit:

\begin{equation}  \label{lambdarrow}
\lim_{M\to\infty}\lambda=\frac{1}{2}\ln 3
\end{equation}
already for $M>10$~(fig.~4). This is in agreement with the
previous finding in eq.~(\ref{lappurerw}) obtained for pure random
walk case.

The diffusion coefficient may be calculated using the second term
of the cumulant expansion of the one-step Lyapunov
exponent~\cite{ZillmerPhD}:
\begin{equation}
D = \lim_{n\rightarrow \infty} \left<\left[\lambda_n-
\left<\lambda_n\right> \right]^2 \right> \label{diffdet}
\end{equation}
Equation (\ref{diffdet}) for our model yields:
\begin{equation}
D=\frac{1}{4}\Biggr[ \ln^2\frac{3+2^{1-M}}{\Delta_M} +
\ln^2\frac{3-2^{1-M}}{\Delta_M} + \ln^2\frac{1+2^{1-M}}{\Delta_M}
+ \ln^2\frac{1-2^{1-M}}{\Delta_M} \Biggr],
\end{equation}
where $\Delta_M=\sqrt[4]{(9-2^{2-2M})(1-2^{2-2M})}$.

The above equation in the case of the pure random
walk~($M\to\infty$) yields:
\begin{equation}
D_{\lambda } = \lim_{M\rightarrow \infty }D=\frac{1}{4}\ln ^{2}3
\label{Dlambdarrow}
\end{equation}
Comparing the above result to eq.~(\ref{diffpurerw}) obtained for
the stochastic model:
\begin{equation}
D_{\lambda} - D_s \cong 0.01 . \label{Differ}
\end{equation}
Thus, we obtained a value which is close to the stochastic
diffusion coefficient (\ref{diffpurerw}). By comparing the basic
statistical properties of stochastically and deterministically
generated series we concluded earlier that the series are
statistically equivalent. Eq.~(\ref{Differ}) may indicate that
although the diffusion coefficient is only a statistical measure
(as in eq.~(\ref{diff})), it may be slightly sensitive to the
dynamical genesis of the system. Still, it would be very difficult
to use such a method even numerically, as usually the statistical
errors are greater than the difference found in
eq.~(\ref{Differ}).

\section{Summary}

We studied the properties of the model defined by eq.~(\ref{eq3})
when the noise generation term is either a stochastic process or a
deterministic process. We compared the time series obtained from
the above implementations of the model by use of DFA. The
conclusion is that the two versions of the model are
indistinguishable using such statistical methods. The advantage of
this observation is that we may calculate the Lyapunov exponent
for the model defined by eq.~(\ref{eq3}). We obtain an equation
relating the DFA exponents (a statistical measure) with the
dynamical Lyapunov exponent for such models. On the other hand,
typical statistical properties can be calculated in two ways:
using statistical analysis or using equations of deterministic
non-equilibrium mechanics. As an example, we calculated the
diffusion coefficient~(\ref{Dlambdarrow}) and obtained a value
which is very close to the stochastic diffusion
coefficient~(\ref{diffpurerw}).

The DFA software used in the calculations, is generously available
from the authors of the method at the Physionet Database World
Wide Web site~(\emph{http://www.physionet.org}~\cite{DFAsr}).

\bibliographystyle{unsrt}

\newpage{}

\section*{Figure captions}
\noindent FIG. 1. In detrended fluctuation analysis the integrated
series $y(k)$ is linearly approximated in each window of size $n$.
Here, two cases of approximation are shown on a fragment of a test
series: for $n=500$ (continuous lines) and $n=2000$ (dashed
lines).
\bigskip\newline
FIG. 2. DFA plots for time series from the model defined by eq.~(\ref{eq3})
at different parameter $M$ values. The length of the time series were $N =
10^5$. \textbf{a)}: $M=1$, this is equivalent to uncorrelated noise series
\textbf{b)}: $M=2$. \textbf{c)}: $M=5$, a crossover occurs - we witness two
slopes. \textbf{d)}: $M=100$, the crossover is shifted in respect to $M=5$.
For window sizes $n$ smaller than the crossover point $n_c$ the slope value
is near to 1.5 (corresponding to Brownian noise), for window sizes larger
than $n_c$ it approaches 0.5 (white noise).
\bigskip\newline
FIG. 3. The equation (\ref{logM}) obtained from numerical simulations.
\bigskip\newline
FIG. 4. The dependence of the Lyapunov exponent $\lambda$ and its
variance on the parameter $M$.

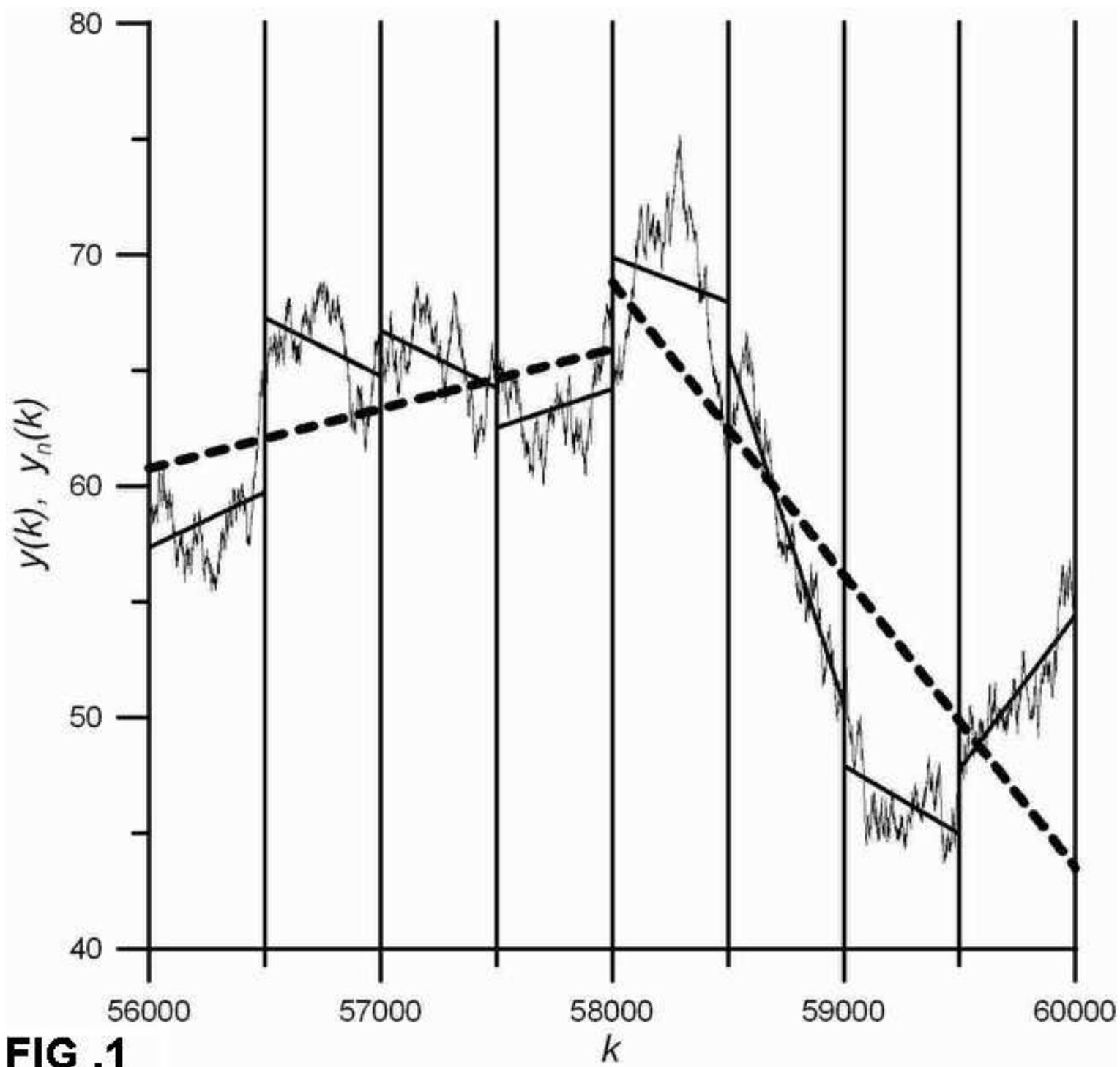
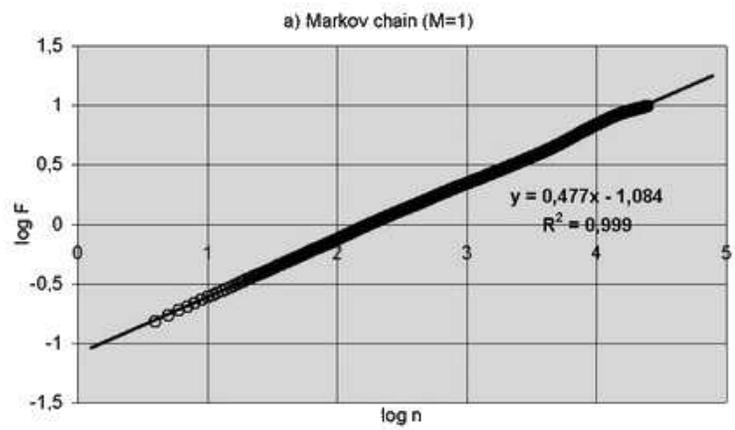
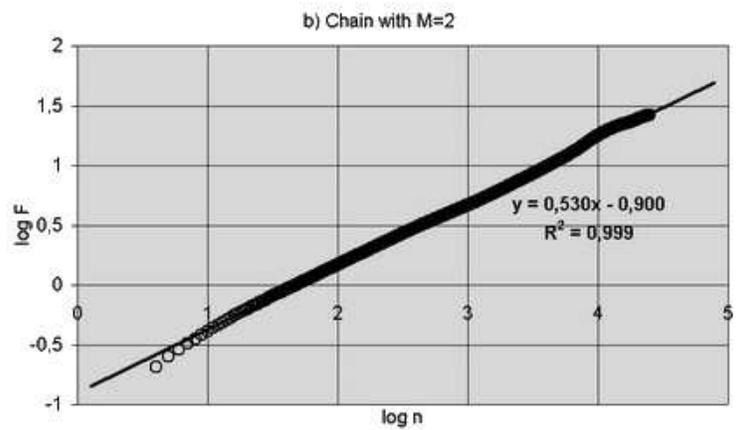
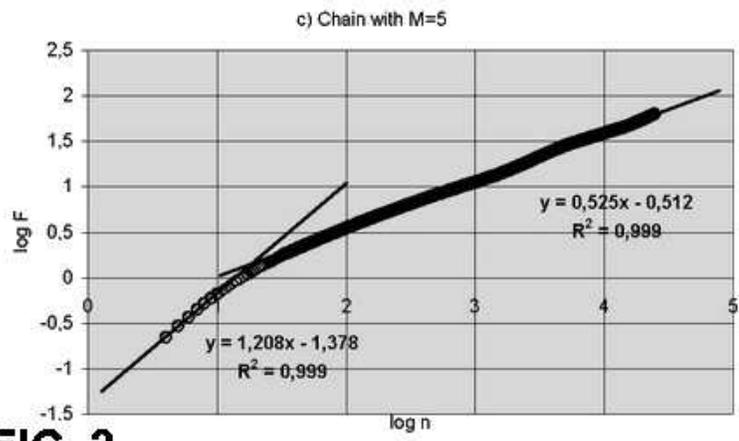
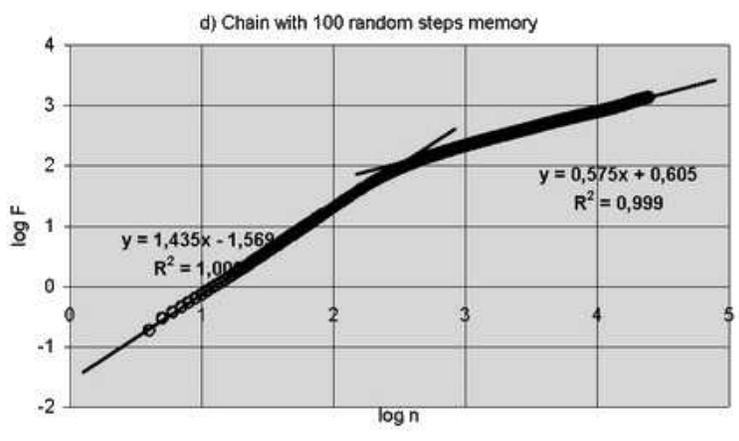
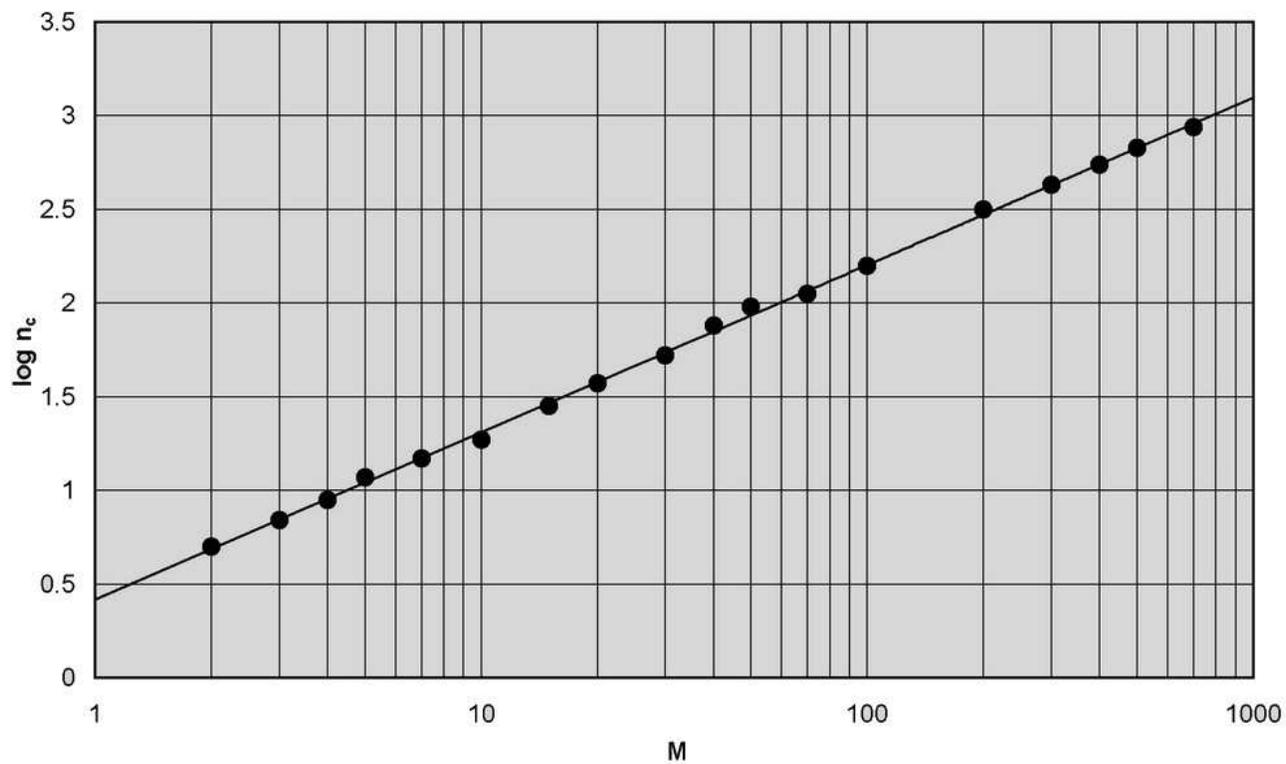
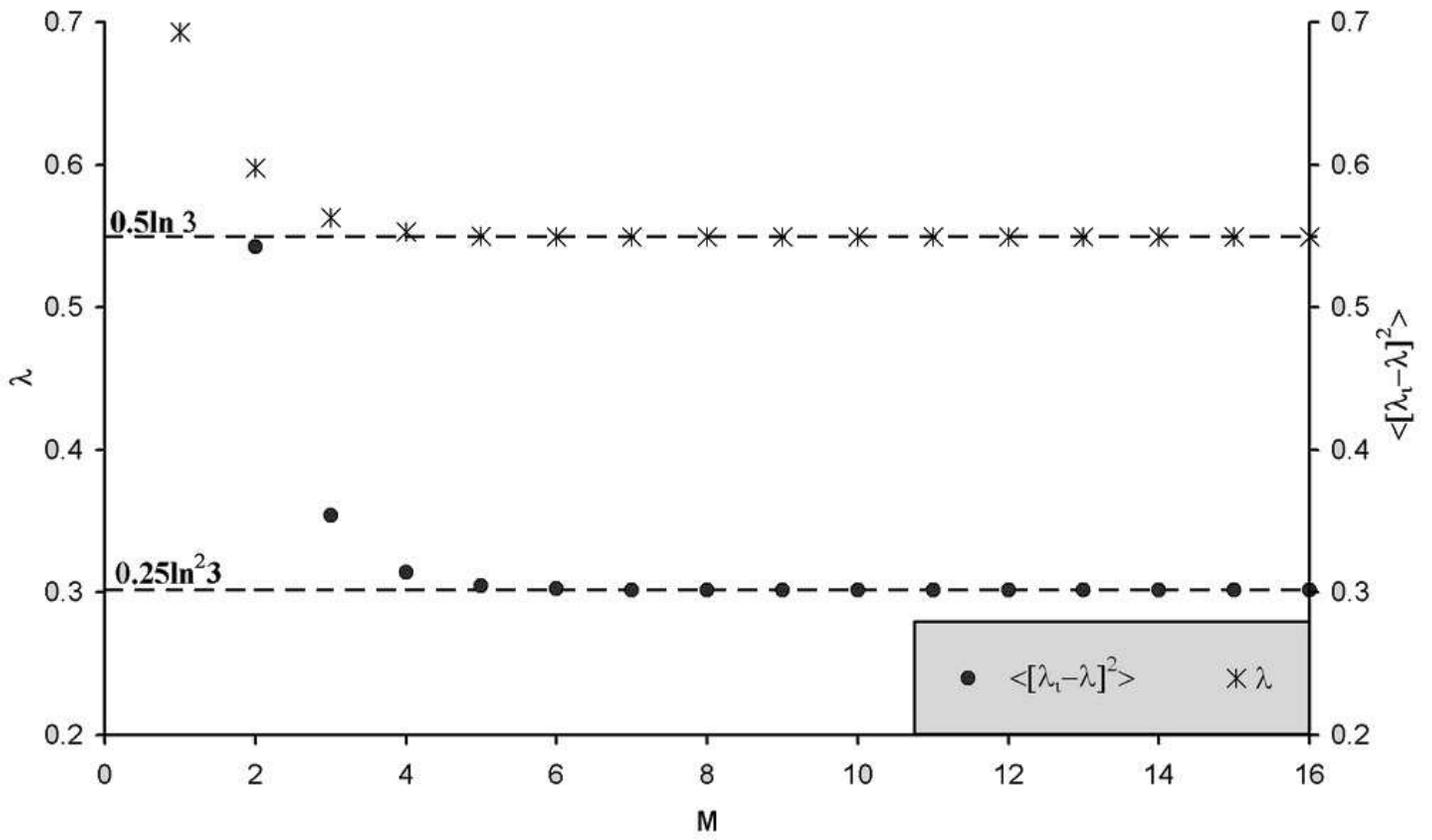
\end{document}